\documentclass[12pt, a4paper]{article}
\usepackage[utf8]{inputenc}
\usepackage{amsmath, amssymb, graphicx}
\usepackage{bm}
\usepackage[margin=1in]{geometry}
\usepackage{setspace}
\usepackage[colorlinks]{hyperref}
\usepackage{doi}
\usepackage{xcolor}
\usepackage{caption}
\usepackage{subcaption}

\usepackage[natbib=true, giveninits=true, style=chicago-authordate-trad, maxnames=2, maxbibnames=99]{biblatex}
\bibliography{references}

\graphicspath{{figs/}}

\title{\vspace*{-2em} \textbf{On the joint volatility dynamics in\\[-.5em]international dairy commodity markets}}
\author{Anthony~N.~Rezitis\footnote{Corresponding author. Professor at the Department of Agricultural Economics and Rural Development, Agricultural University of Athens, Iera Odos 75, 118 55 Athens, Greece. Phone: +30 210 5294735. Email: \href{mailto:arezitis@aua.gr}{arezitis@aua.gr}.}\and Gregor Kastner\footnote{Professor at the Department of Statistics, University of Klagenfurt, Austria. Email: \href{mailto:gregor.kastner@aau.at}{gregor.kastner@aau.at}. Gregor Kastner acknowledges funding from the Austrian Science Fund (FWF) for the project ``High-dimensional statistical learning: New methods to advance economic and sustainability policies'' (ZK 35), jointly carried out by the University of Klagenfurt, WU Vienna University of Economics and Business, Paris Lodron University Salzburg, TU Wien, and the Austrian Institute of Economic Research (WIFO).}}
\date{\today}
\begin{document}

\doublespacing

\maketitle

\vspace*{-2em}

\begin{abstract} \noindent
The present study investigates the price (co)volatility of four dairy commodities -- skim milk powder, whole milk powder, butter and cheddar cheese -- in three major dairy markets.
It uses a multivariate factor stochastic volatility model for estimating the time-varying covariance and correlation matrices by imposing a low-dimensional latent dynamic factor structure.
The empirical results support four factors representing the European Union and Oceania dairy sectors as well as the milk powder markets.
Factor volatilities and marginal posterior volatilities of each dairy commodity increase after the 2006/07 global (food) crisis, which also coincides with the free trade agreements enacted from 2007 onwards and EU and US liberalization policy changes.
The model-implied correlation matrices show increasing dependence during the second half of 2006, throughout the first half of 2007, as well as during 2008 and 2014, which can be attributed to various regional agricultural dairy policies.
Furthermore, in-sample value at risk measures (VaRs and CoVaRs) are provided for each dairy commodity under consideration. 
\end{abstract}



\newpage
\doublespacing

\section{Introduction}
Over the past two decades, the international dairy commodity market has experienced episodic periods of extreme price volatility combined with periods of comparative price stability. In particular since 2007, regional dairy markets in the EU, US, and Oceania have become interrelated and dairy has evolved into a global market \citep{fousekis2016price, newton2016price, fousekis2017price}. Dairy prices are inherently volatile due to the perishable nature of milk, seasonal production, and a combination of an inelastic demand with unanticipated variation in supply caused by factors such as weather and disease. Thus, even small shifts in supply can cause very large changes in price. Furthermore, price volatility is affected by bilateral and multilateral trade agreements and government dairy policy changes enacted around 2007/08. Dairy price volatility in the world market has an impact on virtually all national markets. The size and the speed of this impact, however, may vary \citep{IFCN2017global}.

Rapid dairy price rises and falls since 2007 have increased interest in dairy price volatility, and nowadays price volatility constitutes an important area for empirical research in the agricultural economics literature \citep[e.g.,][]{oconnor2011empirical, briggs2011fluctuations, bergmann2016analysis}. Price volatility is a directionless measure of the extent of the variability of a price. Moderate price variation is desirable for providing price signals that reflect changing market conditions and thus facilitate efficient use of resources. However, extreme price volatility is undesirable with many negative consequences. For example, extremely high prices cause product substitution which might eventually become irreversible, whereas extremely low prices result in financial losses which might result in insolvency.

Changes in supply conditions in significant milk-producing and -exporting regions (such as the EU, the US, and New Zealand) and demand conditions in the main importing regions (such as China, Russia, and Southeast Asia) have impacted dairy commodity prices over the last 20 years. In 2007, world dairy prices increased substantially for skim milk (SMP) and whole milk powder (WMP), respectively, and for the three regions under scrutiny, i.e.\ the EU, Oceania, and the US. This was due to factors related to dairy production, consumption, stocks, trade, and government policies. In particular, in 2007, there was a reduction in supply from the main exporting regions combined with an increase in demand from developing economies in Southeast Asia, the Middle East, and North Africa. The reduction in supply was due to adverse weather conditions in Australia, Argentina, and Uruguay and a higher production cost because of increasing fertilizer prices and cattle feed cost \citep{briggs2011fluctuations}. Furthermore, recent changes in US energy policies, such as the 2007 Energy Independence and Security Act and the 2009 American Recovery and Reinvestment Act, have included many provisions for conservation with grants and tax incentives for both renewable and non-renewable energy which have also contributed to worldwide price volatility. In this context, bilateral and multilateral trade agreements as well as government dairy policy changes, enacted around 2007/08, have sought to liberalize the dairy sector by eliminating trade barriers and facilitating trade. Thus, a more globalized dairy market might have exacerbated dairy price volatility where the price range for SMP, WMP, butter, and cheese expanded after 2007/08 and for the three regions under study \citep{oconnor2009measuring, fousekis2016price,newton2016price,fou-vas:pri,fousekis2016spatial,fousekis2017price}.

The present paper explicitly examines co-volatility price linkages in international dairy markets using a multivariate factor stochastic volatility model, while most of the recent literature investigates price co-movements in international dairy markets without explicitly exploring price volatility relations. In other words, the literature's focal point lies on the co-movement (and transmission) of price levels or changes instead of volatility, as opposed to the present paper, in which volatility transmission is modeled and examined explicitly. For example, \citet{fousekis2016price} investigate price linkages in SMP for the three major exporters (the EU, Oceania, and the US) using a nonlinear autoregressive distributed lag model;
\citet{fou-vas:pri} study price co-movements in the main SMP-producing regions (the EU, Oceania, and the US) using wavelet analysis;
\citet{fousekis2016spatial} examine price dependence in the international butter markets using the copula and wavelet methodologies;
\citet{fousekis2017price} assess the integration of the international SMP market by examining the price relations among the three main producing regions (the EU, Oceania, and the US), applying nonparametric and time-varying copulas; and \citet{rezitis2019impact} examine the impact of trade liberalization on butter and WMP price co-movements between the EU, Oceania, and the US using R-vine copulas.

The study by \citet{rezitis2019impact} provides a discussion of regional dairy policies, mainly in the EU and the USA, and a summary of the various trade agreements around the world. 
These are considered as some of the main causes of international dairy price volatility changes, apart from some natural factors, such as the seasonality of milk production, the perishability of the milk commodity, the inelastic demand, diseases, and adverse weather conditions.
More specifically, the EU dairy sector underwent various policy reforms, such as the 2003 Common Agricultural Policy (CAP) that decreased the dairy intervention prices and introduced the Single Payment Scheme (SPS); and the 2008 CAP Health Check that gradually increased the milk quotas until their final abolition in April 2015. Furthermore, the 2008 Farm Act of the US Dairy Product Price Support Program specified support prices for purchased dairy commodities instead of milk prices; and the 2014 Farm Act abolished subsidy on US dairy exports, since US dairy exports were significantly increasing from 2004 onward. Among the various trade agreements are the trade agreement between China and New Zealand in October 2008; the EU and India, Korea, and the ASEAN countries in April 2007; and the North American Free Trade Agreement (NAFTA) in January 2008; and the Central America Free Trade Agreement (CAFTA) in March 2007. It is worth mentioning that in 2008 Fonterra established Global Dairy Trade delivering electronic auctions (twice a month) in internationally traded dairy commodities like WMP and butter allowing sellers from Europe, North America, Oceania, and India to associate with registered international bidders and thus providing an efficient way for price discovery \citep{rezitis2019impact}.

The aim of this article is to analyze the price volatility of four dairy commodities -- skim milk powder, whole milk powder, butter and cheddar cheese -- in three significant regional markets -- the EU, Oceania, and the US -- for the period from January 20, 2001 to May 13, 2017. These dairy commodities determine international dairy prices as the regions included constitute the leading players of the international dairy market. Our study contributes to the existing literature because -- to the best of our knowledge -- it is the first to employ a multivariate factor SV model, whereas previous studies mainly used multivariate generalized autoregressive conditionally heteroskedastic (GARCH) models. There are several advantages to using the multivariate factor SV approach. First, it reduces the high-dimensional data (i.e.\ 12 international dairy prices) to a lower dimensionality by estimating a small number of latent factors (i.e.\ 4 factors), which are themselves able to exhibit heteroskedasticity.
For this reason, the present paper is able to analyze more dairy products than previous studies \citep[e.g.,][among others]{fousekis2017price, rezitis2019impact}.
Second, these factors can be interpreted as the main driving forces of the international dairy market, being allowed to display volatility clustering and time-varying volatilities. Third, the SV factor model is simple, flexible and robust. Last, the present study may be of interest to dairy industry participants, researchers and policy-makers because it provides a visualization of price volatilities of all dairy commodities under investigation; it shows the proportion of variance of each dairy price which is explained by the common factors; it provides cross-volatility correlations between dairy prices; and it supplies model-implied value at risk (VaR and CoVaR) measures at 5\% and 1\% levels for each dairy commodity.

The GARCH class of models embodies a prominent way of modeling the evolution of volatility deterministically \citep{engle1982autoregressive, bollerslev1986generalized}. Alternatively, it is often appropriate to treat volatility itself as a random quantity and consequently model it stochastically. This can be achieved via a state-space model where the logarithms of the squared volatilities (i.e.\ the latent states) follow an autoregressive (AR) process of order one \citep{taylor1982financial}. The approach later became known as the SV model and through time it has been investigated in numerous studies. Early seminal works in this regard include \citet{jacquier1994bayesian} and \citet{kim1998stochastic}. For a recent empirical comparison of GARCH and SV in the context of modeling agricultural commodity prices, see \citet{yang2018modeling}.

In the multivariate case, the GARCH framework presents a large number of parameters, which are difficult to estimate because of the complicated constraints imposed on those parameters. Simpler versions of multivariate GARCH, such as \citet{bollerslev1990modelling}, assume that the conditional correlations between the series are constant, and thus are not able to model certain complexities of the data properly. On the other hand, multivariate SV models along the lines of \citet{harvey1994multivariate}, \citet{jacquier1994bayesian}, \citet{ghysels1996stochastic}, \citet{kim1998stochastic}, \citet{aguilar2000bayesian}, or \citet{chib2006analysis}, have oftentimes proven to adequately capture multivariate volatility dynamics without being over-parameterized.

Previous studies in commodity price volatility have mainly used the GARCH approach in examining volatility relations and persistence rather than the SV approach of the present study. For example, the study by
\citet{buguk2003price} tested univariate volatility spillovers for prices in the US catfish supply chain;
\citet{apergis2003agricultural} investigated agricultural price volatility spillover effects in Greece;
\citet{Yang2001agricultural} studied agricultural liberalization policy and commodity price volatility;
\citet{gilbert2010food} discussed food price volatility with a special focus on rice;
\citet{jacks2011commodity} investigated commodity price volatility and world market integration;
\citet{serra2013price} studied US corn price fluctuations;
\citet{ji2012how} analyzed the effect of oil price volatility on non-energy commodity markets;
\citet{gardebroek2013do} examined volatility transmission between US oil, ethanol, and corn markets;
\citet{wang2014oil} investigated the responses of agricultural commodity prices to oil price changes;
\citet{cabrera2016volatility} investigated linkages between energy and agricultural commodity prices;
\citet{hernandez2014how} examined the dynamics of volatility across major global exchanges for corn, wheat, and soybeans in the USA, Europe, and Asia; \citet{li2017regimedep} investigated regime-dependent agricultural commodity price volatilities; and \citet{bohl2019speculation} studied the impact of long--short speculators on the volatility of agricultural commodity futures prices, among others. 

The main advantage of the multivariate factor SV model used in the present article is its parsimony. Because variances and covariances are determined by a low-dimensional common factor with the components following independent SV models, even higher-dimensional covariances can be modeled efficiently and without relying on too many parameters.
The factor SV model achieves a combination of simplicity, flexibility, and robustness. Simplicity is obtained by reducing the potentially high-dimensional observation space to a lower-dimensional orthogonal latent factor space, in a similar way as in the classic factor model case. Flexibility is achieved by allowing these factors to display volatility clustering. Robustness is reached because idiosyncratic deviations are themselves stochastic volatility processes, thus permitting for the degree of volatility co-movement to be time-varying. 
For the study at hand, we use a recently developed method for Bayesian inference \citep{kastner2017efficient}, which combines an efficient method for estimating univariate SV models \citep{kastner2014ancillarity} with a standard Gibbs sampler for regression problems. This method also uses interweaving strategies \citep{yu2011to} and thus it achieves a quicker convergence and better mixing of the Markov chain Monte Carlo (MCMC) method.

The remainder of the article is structured as follows. In Section~\ref{sec:model}, the factor SV model is introduced, some of its key properties are discussed, and the Bayesian estimation methodology is described. Section~\ref{sec:data} focuses on the data and discusses some important descriptive statistics. The empirical findings are reported in Section~\ref{sec:res}. Finally, Section~\ref{sec:con} concludes.

\section{Model specification and estimation}
\label{sec:model}

We first introduce the factor SV model in Section~\ref{sec:fsv}, followed by Section~\ref{sec:priors} which discusses the prior choices. Finally, Section~\ref{sec:estimation} briefly discusses the estimation methodology.

\subsection{The factor stochastic volatility model}
\label{sec:fsv}

The two main ingredients that comprise the dynamic covariance model used in this study are, on the one hand, the univariate SV model, and, on the other hand, the factor model. Observing $m$ time series $y_{it}$, where $i = 1,\dots,m$ and $t = 1,\dots,T$, we first fix the number of latent factors $r$ and consequently specify $m + r$ conditionally independent univariate SV models. More concretely, for $i = 1,\dots m+r$, let
\begin{equation}
h_{it} = \mu_i - \phi_i(h_{i, t-1} - \mu_i) + \sigma_i \eta_{it}, \quad \eta_{it} \sim \mathcal{N}(0, 1),
\label{ar}
\end{equation}
denote latent AR(1) processes driving the time dynamics of both factor log variances as well as residual (\emph{idiosyncratic}) log variances. The parameter $\mu_i \in \mathbb{R}$ is the \emph{level}, $\phi_i \in (-1, 1)$ the \emph{persistence}, and $\sigma_i \in \mathbb{R}^+$ the \emph{volatility} of log variance $\bm h_i = (h_{i0}, h_{i1}, \dots, h_{iT})$. Furthermore, stack these -- nonlinearly, to guarantee positivity -- into the following two diagonal matrices
\begin{equation}
\bm{U}_t = \text{Diag}(e^{h_{1t}},\dots,e^{h_{mt}}), \quad \bm{V}_t = \text{Diag}(e^{h_{m+1,t}},\dots,e^{h_{m+r,t}}).
\end{equation}
Finally, define the $r$-dimensional latent factors
\begin{equation}
\bm{f}_t \sim \mathcal{N}_r(\bm{0},\bm{V}_t),
\end{equation}
and combine these with factor loadings $\bm{\Lambda} \in \mathbb{R}^{m \times r}$ to model the data through an $m$-variate conditionally Gaussian distribution,
\begin{equation}
\bm{y}_t = \bm{\Lambda}\bm{f}_t + \bm{\epsilon}_t, \quad \bm{\epsilon}_t \sim \mathcal{N}_m(\bm{0}, \bm{U}_t).
\label{model}
\end{equation}
The initial distribution of the latent log variances is given through $h_{i0} \sim \mathcal{N}(\mu_i, \sigma^2_i/(1-\phi_i^2))$, meaning that each $h_{i0}$ is assumed to come from the stationary distribution of (\ref{ar}). To identify the scale of the factors we set $\mu_i = 0$ for $i > m$ \citep[see][for details]{kastner2017efficient}.

Some comments are in order. First, the model defined in (\ref{ar}) to (\ref{model}) implies that the observations -- after integrating out the latent factors -- have mean zero and covariance matrix
\begin{equation}
\bm{\Sigma}_t = \bm{\Lambda}\bm{V}_t\bm{\Lambda}' + \bm{U}_t.
\label{cov}
\end{equation}
As $\bm{V}_t$ and $\bm{U}_t$ are diagonal matrices, the dependence structure of the data is solely driven through the factor volatilities in $\bm{V}_t$ and the factor loadings in $\bm{\Lambda}$, whereas the idiosyncratic volatilities in $\bm{U}_t$ model (temporary) deviations therof. These diagonality assumptions are essential in the context of factor models to guarantee sparsity of the covariance matrix representation (\ref{cov}) which can be rewritten in terms of dynamic loadings $\bm{\Sigma}_t = \bm{\Lambda}_t\bm{\Lambda}'_t + \bm{U}_t$ with $\bm{\Lambda}_t := \bm{\Lambda}\bm{V}_t^{1/2}$ and imply that the factors are a priori uncorrelated.\footnote{For possible generalizations, we point toward \citet{bai-ng:det} for an exposition to approximate factor models where $\bm{U}_t$ need not be diagonal and to \citet{zhou2014bayesian} for a model featuring correlated factors.}
 
Second, note that the only observables in this model are the data $\bm{y}$, whereas all other latent variables and parameters are to be estimated. Third, the ordering and the sign of the factors is not likelihood-identified, which means that these cannot be inferred from the data. In what follows, we chose an arbitrary ordering of the factors and assume that the largest loading on each factor is positive.\footnote{Note that these assumptions do not restrict the likelihood nor do they matter for the estimated and predicted (co-)volatilities. They are, however, relevant for interpreting the latent factors and their loadings.}

\subsection{Prior specification}
\label{sec:priors}

Our approach to estimation is Bayesian, which means that prior distributions have to be specified for all parameters. Concerning the level of the log variances, we use a standard uninformative Gaussian prior $\mu_i \sim \mathcal{N}(0, 100)$ for $i = 1,\dots,m$. In order to bound the persistence to the interval $(-1, 1)$, we use a beta prior on an affine transformation of $\phi_i$, i.e., $(\phi_i + 1)/2 \sim \mathcal{B}(20, 1.5)$ for $i = 1,\dots,m+r$. This specific choice shifts prior mass close to 1 which translates to persistent log variances \citep[cf.][]{kim1998stochastic}. For the volatility of log variance we use a gamma prior, $\sigma_i^2 \sim \mathcal{G}(1/2, 1/2)$ for $i = 1,\dots,m+r$, which is equivalent to stipulating a standard normal prior on $\pm\sigma_i$. This choice -- as opposed to the traditionally more commonly employed inverse gamma prior -- does not bound away $\sigma_i$ from zero \citep[see][for a more elaborate discussion of this issue]{fruhwirth2010stochastic, kastner2014ancillarity}. We note that for the analysis at hand, this elicitation exhibits only minor inferential influence and the posterior proves to be quite stable when the prior hyperparameters are varied. The priors for the parameters are visualized in Figure~\ref{fig:parameters} in Section~\ref{sec:volload}. Finally, for the factor loadings matrix, we employ standard normal priors independently for each element of $\bm{\Lambda}$. We also experimented with alternative (shrinkage) prior specifications for $\bm{\Lambda}$ such as those discussed in \citet{kastner2019sparse} leading to qualitatively similar results. Overall, we note that the results in Section~\ref{sec:res} appear to be quite robust to the prior hyperparameter choices.

\subsection{Posterior inference}
\label{sec:estimation}

Estimation itself is carried out via the R \citep{rlanguage} package factorstochvol \citep{hos-kas:mod}. This software collection implements the MCMC algorithm of \citet{kastner2017efficient} which uses several ancillarity-sufficiency interweaving strategies \citep[ASIS,][]{yu2011to} to ensure convergence and appropriate mixing of the Markov chain. For posterior inference, we use $100\,000$ MCMC draws after a burn-in of $50\,000$ (which we discard). To avoid excessive computer memory usage, we furthermore thin the retained draws with a factor of $100$. To conclude, the posterior samples are checked for convergence by visually investigating traceplots of the Markov chains and computing effective sample sizes for all quantities of interest through the R package coda \citep{plummer2006coda}.

Concerning order and sign identification of the factors and their loadings, we choose not to impose any restrictions such as the commonly employed lower-unitriangular factor loadings matrix \citep[e.g.,][]{aguilar2000bayesian, chib2006analysis, han2006asset, zhou2014bayesian} but instead ``let the data speak''. To this end, we run the MCMC chain on an unrestricted loadings space and identify signs through post-processing where we require the highest loading on each factor to have positive sign.

\section{Data}
\label{sec:data}

\begin{figure}[p]
    \centering
    \includegraphics[width=\textwidth]{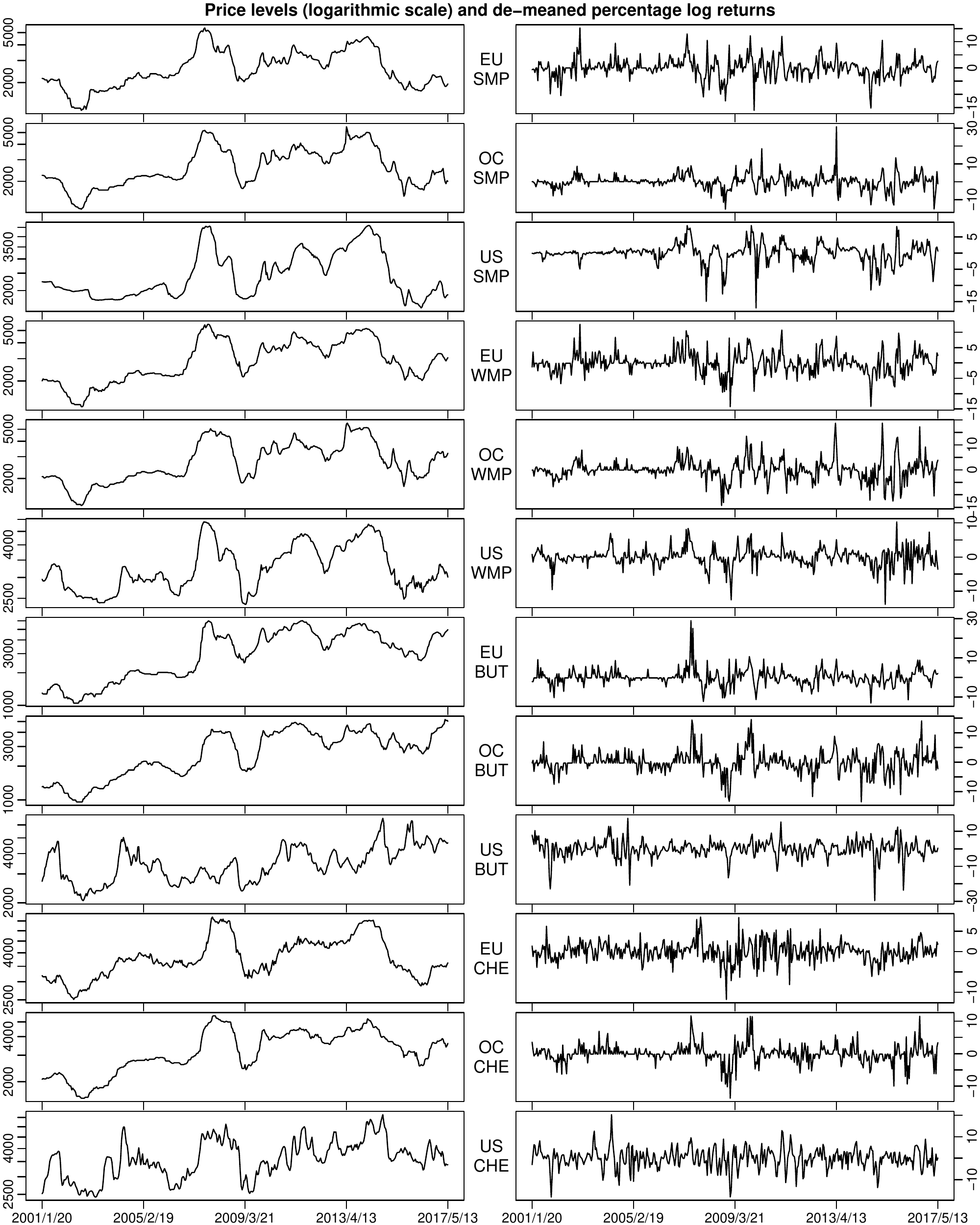}
    \caption{Bi-weekly prices and returns of skim milk powder (SMP, top 3 panels), whole milk powder (WMP, panels 4--6), butter (BUT, panels 7--9), and cheddar cheese (CHE, bottom 3 panels) for the European Union (EU), Oceania (OC), and the United States (US).}
    \label{fig:data}
\end{figure}

Price data on four dairy commodities, that is, skim milk powder, whole milk powder, butter, and cheddar cheese, in three regions, namely, the EU, Oceania (New Zealand and Australia), and the US, have been obtained from the databases of the Dairy Marketing and Risk Management Program/University of Wisconsin and are collected from the Dairy Market News of USDA/AMS. Price data are measured in dollars per metric ton and the time period considered ranges from January 6, 2001 to May 13, 2017. The data set contains $m=12$ (all of which are available for this time frame) bi-weekly dairy prices (i.e.\ prices of four dairy commodities in three regions) on $426$ dates. For further analysis, we thus use $T=425$ demeaned log dairy price changes (i.e.\ log returns which have been shifted by a constant to have unconditional mean zero).
Figure~\ref{fig:data} provides plots of prices and log returns.

The four major exporters of dairy products (i.e.\ SMP, WMP, Butter, and Cheese) in the period 2013--2015 are Oceania with a share of about 38\% (i.e.\ New Zealand with about 32\%, and Australia with about 6\%), the European Union (31\%), and the United States (12\%).

\begin{table}[t]
 \centering
 \caption{Dairy commodity exports. Averages 2013--2015 in 1000 tonnes product weight. Source: \citet{OECD2016}}
 \begin{tabular}{rrrrrrrrr}
  \hline
  & \multicolumn{2}{c}{SMP} & \multicolumn{2}{c}{WMP} & \multicolumn{2}{c}{Butter} & \multicolumn{2}{c}{Cheese} \\
  \hline
  & Quantity & \% & Quantity & \% & Quantity & \% & Quantity & \% \\
  \hline
  EU-28 & 0.7069 & 33.59 & 0.4249 & 16.94 & 0.2294 & 24.89 & 1.0775 & 44.14 \\
  Oceania & 0.5960 & 28.32 & 1.4305 & 57.02 & 0.5284 & 57.33 & 0.4542 & 18.61 \\
  USA & 0.5485 & 26.07 & 0.0177 & 0.70 & 0.0595 & 6.45 & 0.3331 & 13.65 \\
  Other & 0.2529 & 12.02 & 0.6356 & 25.34 & 0.1044 & 11.33 & 0.5760 & 23.60 \\
  World & 2.1043 & & 2.5087 & & 0.9217 & & 2.4408 \\
  \hline
 \end{tabular}
 \label{tab:exports}
\end{table}

Based on Table~\ref{tab:exports}, the leading SMP exporters are the EU-28, which accounts for about 33.6\% of the total world exports, Oceania with about 28.3\%, and the US with approximately 26\% of the total exports. SMP is extensively used in the food industries and for recombination with butter oil and other fat products. It is an important product for low-income countries and for countries with insufficient fresh milk production like China. Such countries are going to become important export destinations in the future \citep{thiele2013economic}. The larger world importers of SMP are Southeast Asia, North Africa, China, and Mexico. Trade flows amongst the areas under consideration (i.e.\ EU-28, Oceania, and the US) are negligible and market integration may be attained indirectly through competition in the international SMP market.

In the WMP international market, Oceania has dominated with a share of about 57\%, while the EU-28 accounted for approximately 17\%, and the USA for less than 1\% (Table~\ref{tab:exports}). WMP is an important ingredient in the food industry, providing both milk fat and non-fat milk solids. It is demanded more by countries with strong economic power, like oil-exporting countries. China, the Middle East, North Africa, and Southeast Asia are the major import markets. As in the case of SMP, trade flows are negligible among the EU-28, Oceania, and the US, and market integration may be achieved only through competition in the international WMP market.

In the international butter market, Oceania has been the main exporter, with a market share of above 57\%, followed by Europe and the US with about 25\% and 6.5\%, respectively (Table~\ref{tab:exports}). In the case of butter, the three regions under consideration interact with each other through direct trade because substantial butter volumes are imported in the US from various EU counties (mainly Ireland) and from New Zealand \citep{groves2016us}. Moreover, substantial butter exports take place from New Zealand to the EU due to trading agreements and duty free access for inward processing \citep{fousekis2016spatial}. Furthermore, these regions compete indirectly in the international butter market. Among the major importers are the Middle East, Russia, North Africa, Southeast Asia, and China.

In the worldwide cheese trade, the EU-28 was the main exporter, with about a 44\% share, followed by Oceania with about 19\%, and the US with about 14\% (Table~\ref{tab:exports}). The larger importers of cheese (in value) in 2015 were Russia, the US, and Japan \citep{davis2016assessing}. Significant trade also takes place in countries with a lower level of consumption but with consumers familiar with cheese, such as Latin American, North African, and Middle East countries \citep{thiele2013economic}. Thus, the three regions (i.e.\ the EU-28, Oceania, and the US) interact between each other through direct trade as well as indirect trade in the international cheese market.

Despite the low share of the US in total world exports in the cases of WMP and butter (i.e.\ 0.75\% and 6.46\%, respectively), we are including the US in our analysis because it is expected to play a significant role in the international dairy market in the future due to its bilateral trade with the EU \citep{boulanger2016cumulative}. This is especially so after July 2018, when President Donald Trump and the EU decided to resume talks regarding bilateral trade similar to the Transatlantic Trade and Investment Partnership (TTIP) negotiations, which were halted by the Trump administration initiating a trade conflict with the EU. However, due to the lack of more updated data, we do not consider these changes in the present article.



\section{Results}
\label{sec:res}

In this section, we present the results of the estimation procedure. We begin by discussing the factor loadings and the estimated factor volatilities, alongside the communalities, in Section~\ref{sec:volload}. Thereby, we gain structural insight into the covariance structure of the international dairy commodity market. In Section~\ref{sec:vol}, we discuss the marginal volatilities of each component series and the parameter estimates of the idiosyncratic volatilities. In Section~\ref{sec:cor}, we turn towards examining the model-implied dynamic correlation structure per se. Finally, in Section~\ref{sec:var}, we look at in-sample value at risk measures (VaRs and CoVaRs) based on the model-implied posterior median covariance matrix.  

\subsection{Factor volatilities and loadings}
\label{sec:volload}

\begin{figure}[t]
    \centering
    \includegraphics[width=.495\textwidth,page=1]{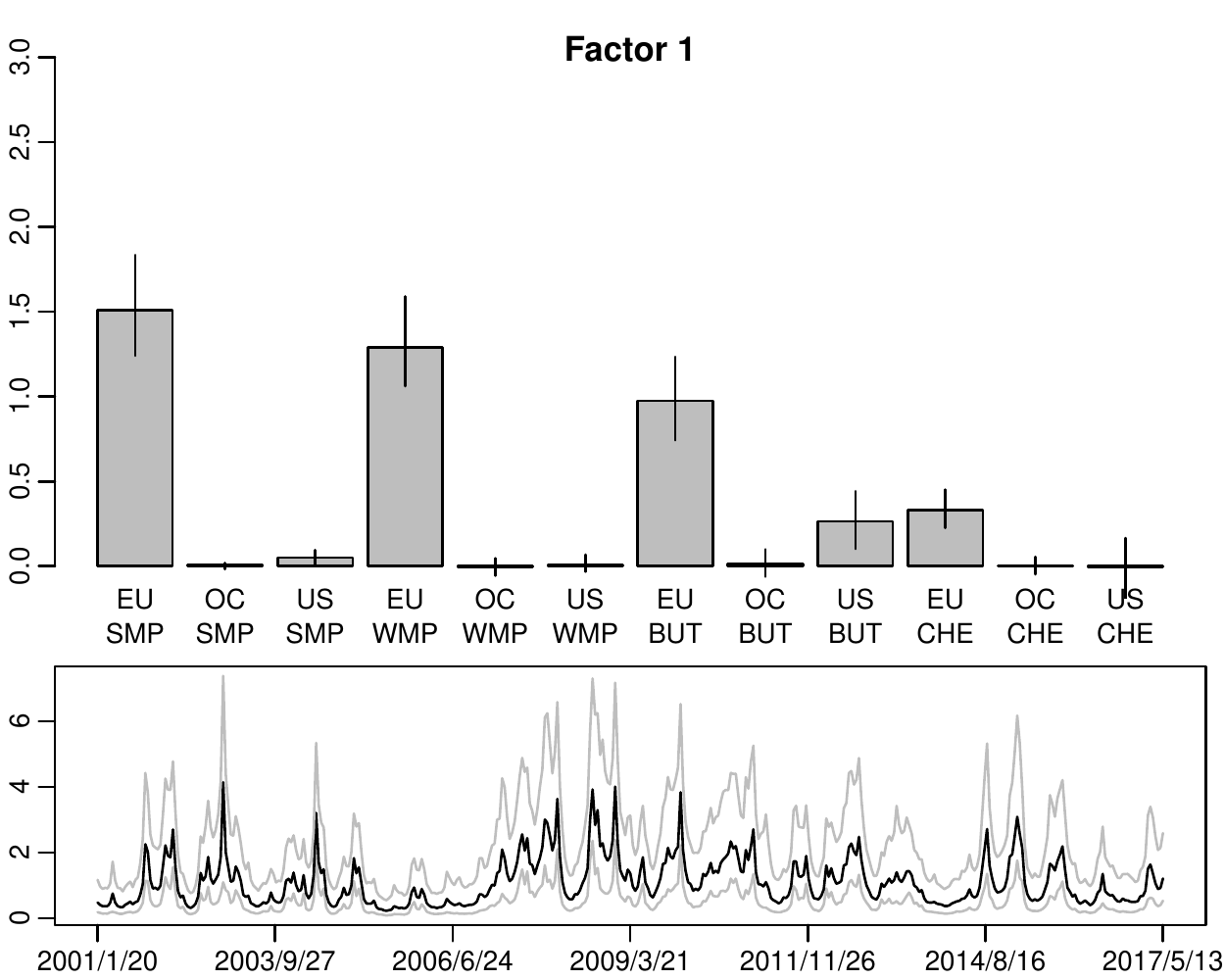}\hfill
    \includegraphics[width=.495\textwidth,page=2]{factors.pdf}\\
    \includegraphics[width=.495\textwidth,page=3]{factors.pdf}\hfill
    \includegraphics[width=.495\textwidth,page=4]{factors.pdf}
    \caption{Loadings (top) and volatilities (bottom) of the four estimated factors (0.1/0.5/0.9 posterior quantiles).}
    \label{fig:factors}
\end{figure}

Factor loadings are presented in Figure~\ref{fig:factors}, alongside the corresponding log variances of the latent factors. The first factor can be interpreted as the EU dairy sector as all dairy EU variables (EU--SMP, EU--WMP, EU--BUT, and EU--CHE) load very highly on this factor. This is supported by the fact that the EU is one of the world's leading dairy exporters. The loadings of the second factor are primarily the Oceania and secondarily the European milk powder variables, i.e.\ OC--SMP, OC--WMP, EU--SMP, and EU--WMP. This finding can be explained by the fact that both Oceania and Europe are among the main world milk powder exporting regions. The third factor can be viewed as the Oceania dairy sector as all Oceania dairy variables (OC--SMP, OC--WMP, OC--BUT, and OC--CHE) load highest on this factor. This finding is also expected because Oceania, like the EU, is one of the driving forces of the world dairy export market. Investigation of the fourth factor indicates that the US should also be considered as an additional force in the milk powder market. This is because the US milk powder variables (US--SMP and US--WMP) load highly on the fourth factor. Note that the other variables that are loaded significantly on the fourth factor are the EU and Oceania milk powder variables (EU--SMP, EU--WMP, OC--SMP, and OC--WMP). It seems that the fourth factor expands the second factor by including the US milk powder variables as additional loadings to those of the second factor. This finding suggests that the US is a significant player in the international SMP market but to a lesser extent than the EU and Oceania. This argument is also supported by \citet{fou-vas:pri, fousekis2017price}, who find that the price integration and linkages have been stronger between the EU and Oceania. The US has been catching up with them since it was a latecomer to the SMP global market.   

One common feature of the latent factors is that their volatilities increased after the 2006/07 period.
This could be attributed to the greater dairy market integration that occurred after 2007 due to the various bilateral and multilateral trade agreements enacted and gradually implemented after this year as well as to the agricultural policy changes in the EU and the US. The more globalized dairy market might have exacerbated the dairy price volatility due to several causes, such as supply and demand shocks, uncertainty about dairy stock levels, adverse weather conditions, climate changes, financial speculation, exchange rate volatility, and oil price volatility, among others.
From the covariance decomposition in Equation \ref{cov}, it follows that higher factor volatilities directly correspond to higher covariances during this period.
This finding is also supported by the previous literature. For example, \citet{fousekis2017price} indicate an increasing degree of price co-movement as well as statistically significant probabilities of joint price crashes and booms for the three main SMP-producing regions (the EU, Oceania, and the US); \citet{fousekis2016spatial} support strong long-run price linkages in the two butter-producing regions (the EU and Oceania); \citet{fousekis2016price} suggest stable long-run price linkages in SMP in the three regions (the EU, Oceania, and the US) but find that the law of one price does not hold; and \citet{rezitis2019impact} find slightly increasing price dependence after the year 2007 in the butter and WMP markets for the EU--Oceania and EU--US pairs, which is attributed to the aftermath of the liberalization changes. However, their results provide evidence that the dairy market is segmented rather than well integrated. 
Furthermore, the first factor, which is related to the EU dairy sector, shows an increasing volatility around 2003, which can be attributed to the 2003 CAP reform,  and high volatility during the subperiod 2007--2010 and around 2014. The second and fourth factors, which are related to milk powder, exhibit similar patterns and clearly an increase in volatility after the period 2006/07. Note that the second factor shows an increase in volatility around 2003, whereas the third factor, which is attributed to the Oceania dairy sector, displays high volatility around the subperiods 2007--2010 and 2014--2016. Note that the first and the third factors show higher volatility and smaller persistence in comparison to the second and the fourth factors.

\begin{figure}[p]
    \centering
    \includegraphics[width=\textwidth, trim=0 0 0 12, clip]{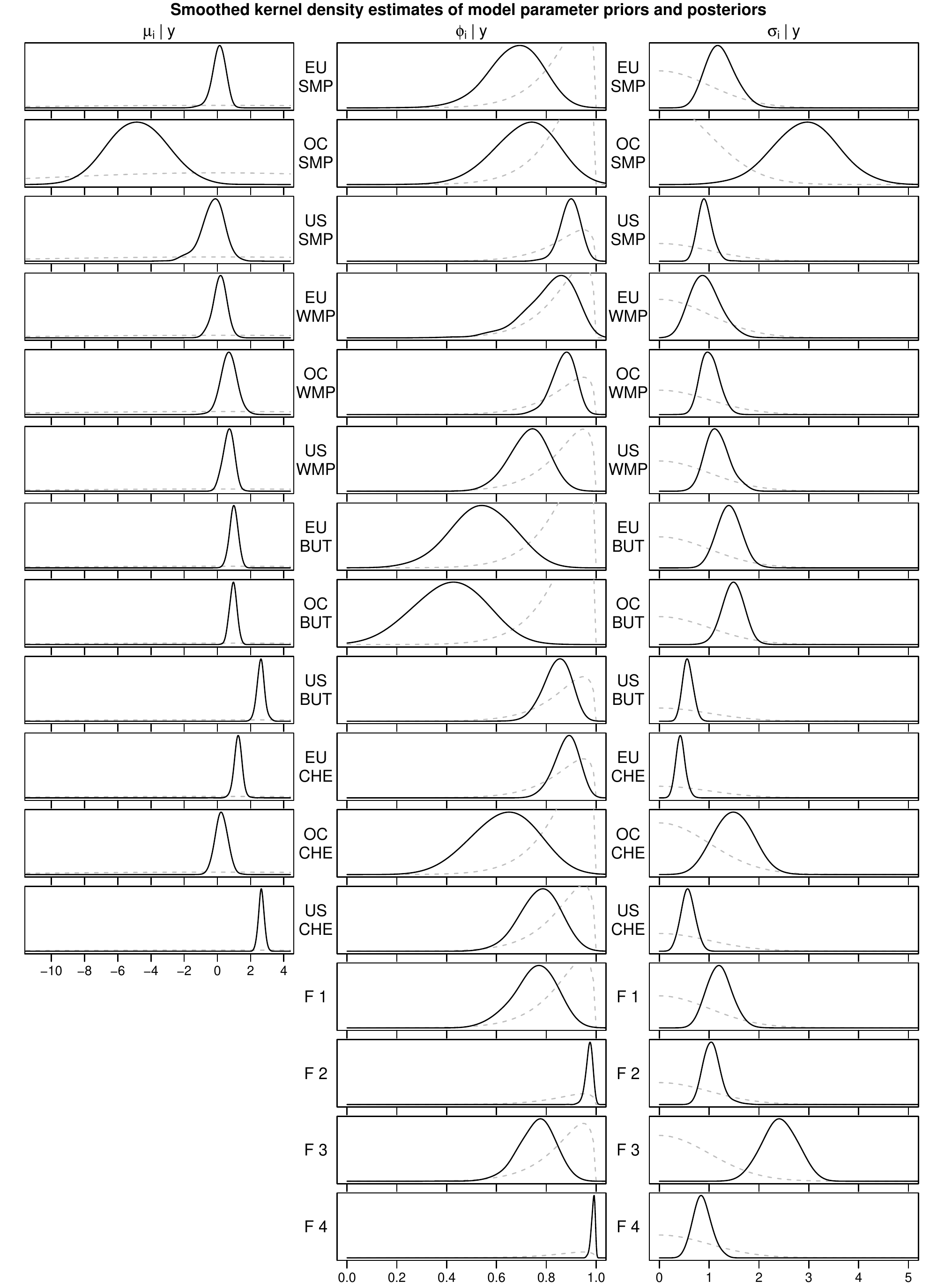}
    \caption{Prior (dashed, gray) and posterior (solid, black) distributions of the model parameters. The first 12 rows correspond to the idiosyncratic volatilities and the bottom 4 rows correspond to the factor volatilities.}
    \label{fig:parameters}
\end{figure}

This finding is corroborated by Figure~\ref{fig:parameters}, which displays the posterior distributions of the model parameters $\mu_{i}$, $\phi_i$, and $\sigma_i$ (i.e. the level, the persistence, and the volatility of log variance, respectively). More specifically, the bottom four rows of Figure~\ref{fig:parameters} correspond to the four factors. Note in particular that the persistence of factor two and factor four are very close to one, whereas the persistence of factor one and three are only around $0.8$. Furthermore, the volatility of the third factor is much higher than the volatility of the rest of the factors, while the volatility of the first factor is a little higher than the volatility of the remaining two factors. 

\begin{figure}[p]
    \centering
    \includegraphics[width=\textwidth, trim=0 0 0 12, clip]{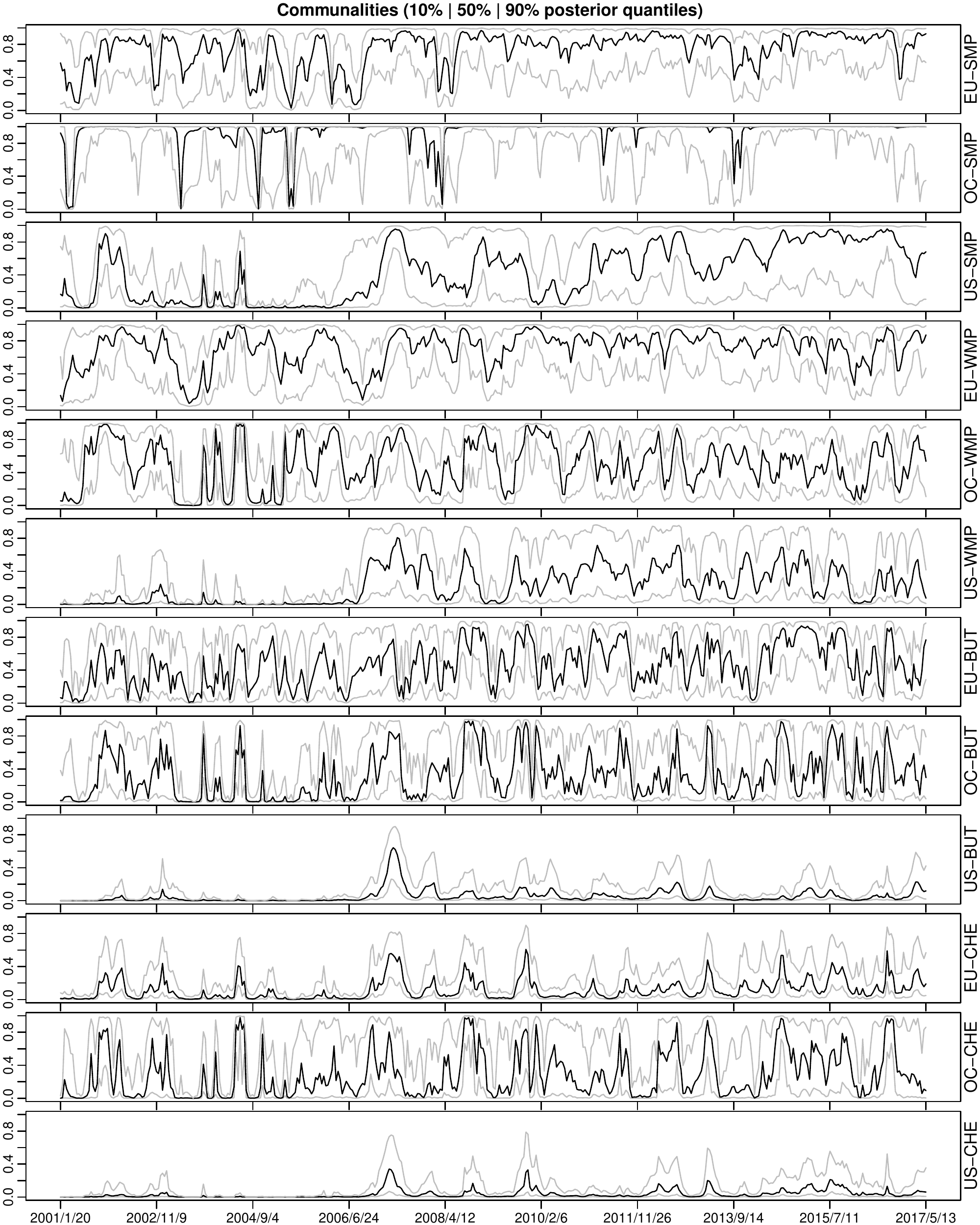}
    \caption{Communalities, i.e.\ the proportions of marginal variance explained through the common factors (0.1/0.5/0.9 posterior quantiles).}
    \label{fig:com}
\end{figure}

The above-mentioned four factors are supported by the communalities, which are presented in Figure~\ref{fig:com}. Communalities show the proportion of marginal variance of each variable which is explained by the common factors. The common factors explain a significant amount of the variance of each one of the variables under consideration except those of US butter and cheese (i.e.\ US--BUT and US--CHE). This is in line with the fact that neither US--BUT nor US--CHE load significantly on any of the four factors of our Factor Stochastic Volatility (FSV) model. It is worth noting, however, that based on Figure~\ref{fig:com} the proportion of the variance of the US whole milk powder explained by common factors is significant after 2006. Note that, based on \citet{fou-vas:pri, fousekis2017price}, the dairy price integration and linkages have been stronger between the EU and Oceania, and this might be because the US entered the international dairy markets much later.  

\subsection{Marginal volatilities}
\label{sec:vol}

\begin{figure}[p]
    \centering
    \includegraphics[width=\textwidth, trim=0 0 0 12, clip]{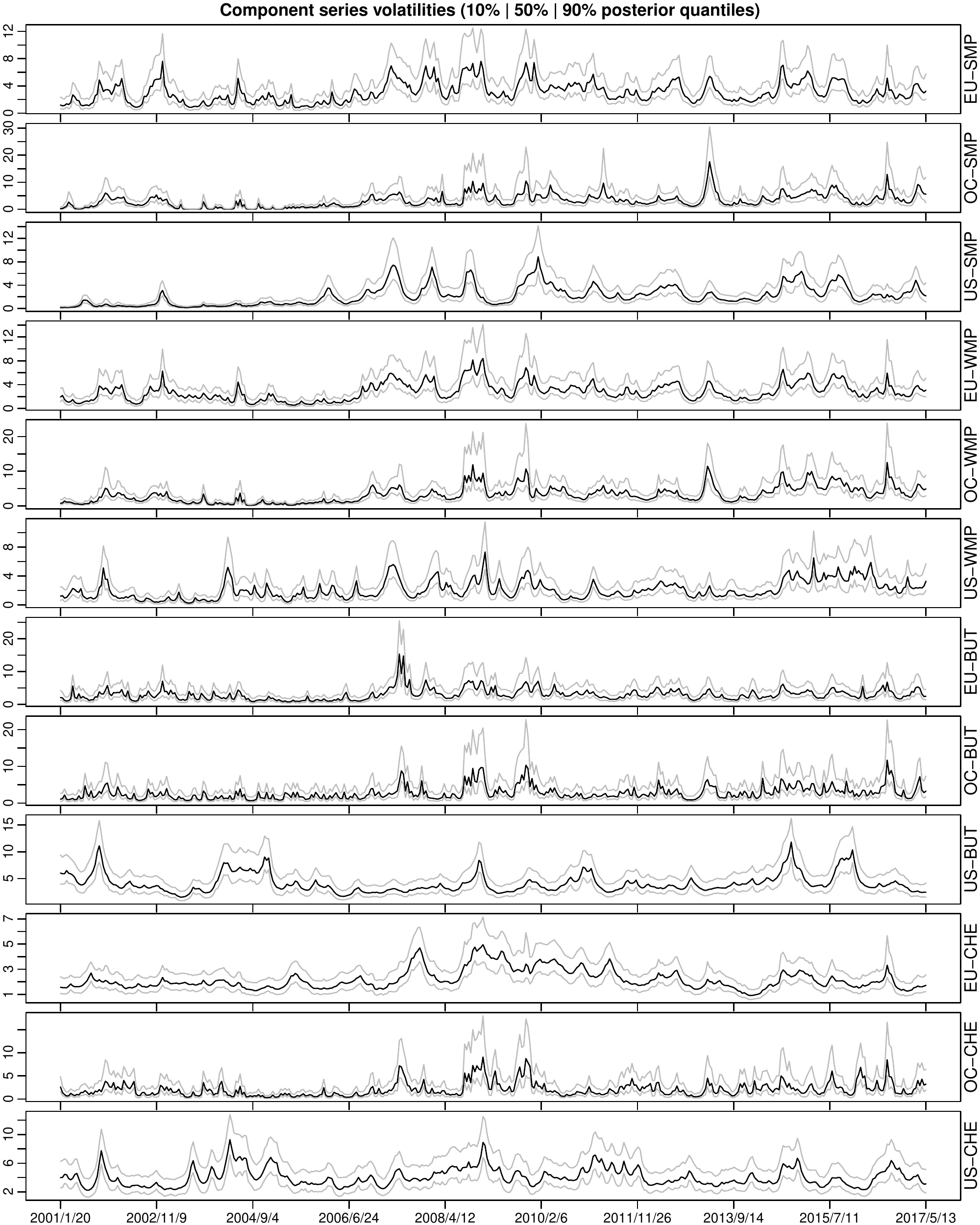}
    \caption{Marginal volatilities (0.1/0.5/0.9 posterior quantiles).}
    \label{fig:vol}
\end{figure}

A thorough visualization of the marginal posterior volatilities of all dairy commodities, presented in Figure~\ref{fig:vol}, indicates that the volatility of each of the dairy commodities increases after the 2006/07 global (food) crisis period. In general, the graphs show that for most of the commodities, volatility rates are higher during the subperiods 2007--2010 and 2014--2016. Moreover, the pattern of marginal volatilities among the dairy commodities in Figure~\ref{fig:vol} shows strong heteroskedasticity as well as considerable co-movement, providing some empirical support for multivariate modeling through common latent factors.  

An examination of the top twelve rows of Figure~\ref{fig:parameters}, showing the posterior distributions of the level ($\mu_{i}$), persistence ($\phi_i$), and volatility ($\sigma_i$) of the idiosyncratic log variance of the components corresponding to each of the dairy commodities under consideration, indicates that EU--CHE, OC--WMP, and US--SMP exhibit the highest persistence amongst all idiosyncratic volatilities. Dairy commodities with the highest idiosyncratic volatility of log variance $\sigma_i$, $i=1,\dots,m$, are OC--SMP, OC--BUT, OC--CHE, and EU--BUT. This finding is in line with the observation that factor three, representing Oceania’s dairy sector, shows a higher volatility of log variance than the others.   

\subsection{Posterior correlations}
\label{sec:cor}



\begin{figure}[tp]
     \centering
     \begin{subfigure}[b]{0.193\textwidth}
         \centering
         \includegraphics[width=\textwidth, page=5]{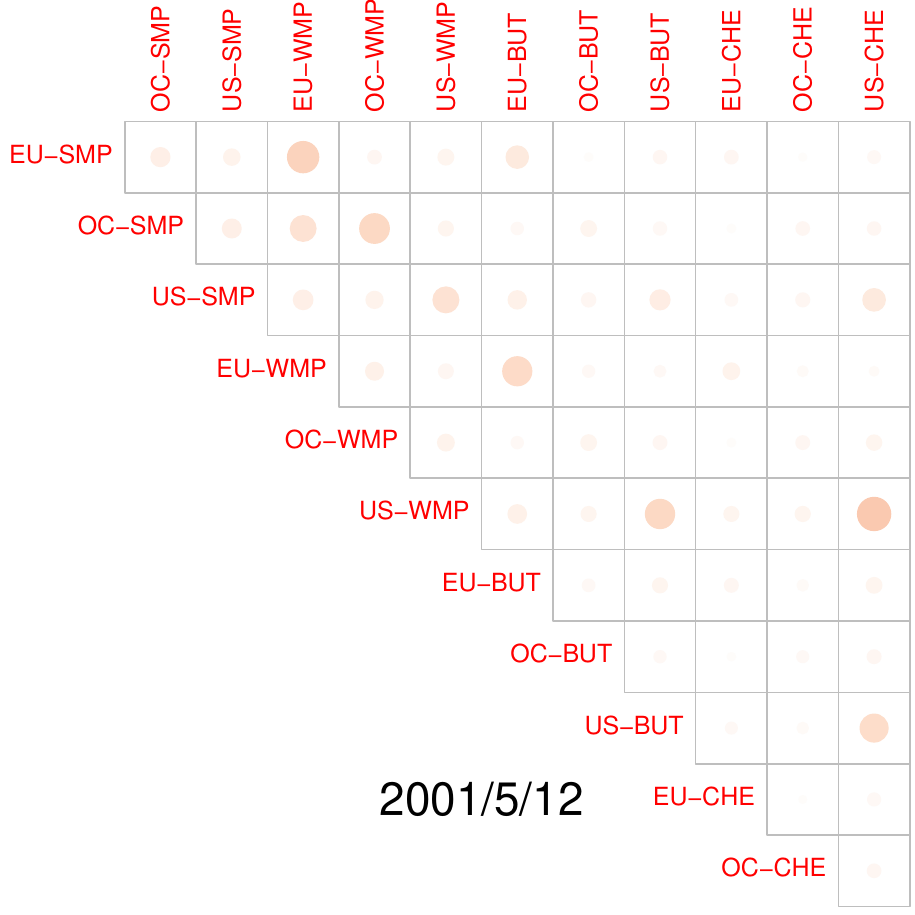}
         \caption{}
         \label{fig:a}
     \end{subfigure}
     \begin{subfigure}[b]{0.193\textwidth}
         \centering
         \includegraphics[width=\textwidth, page=47]{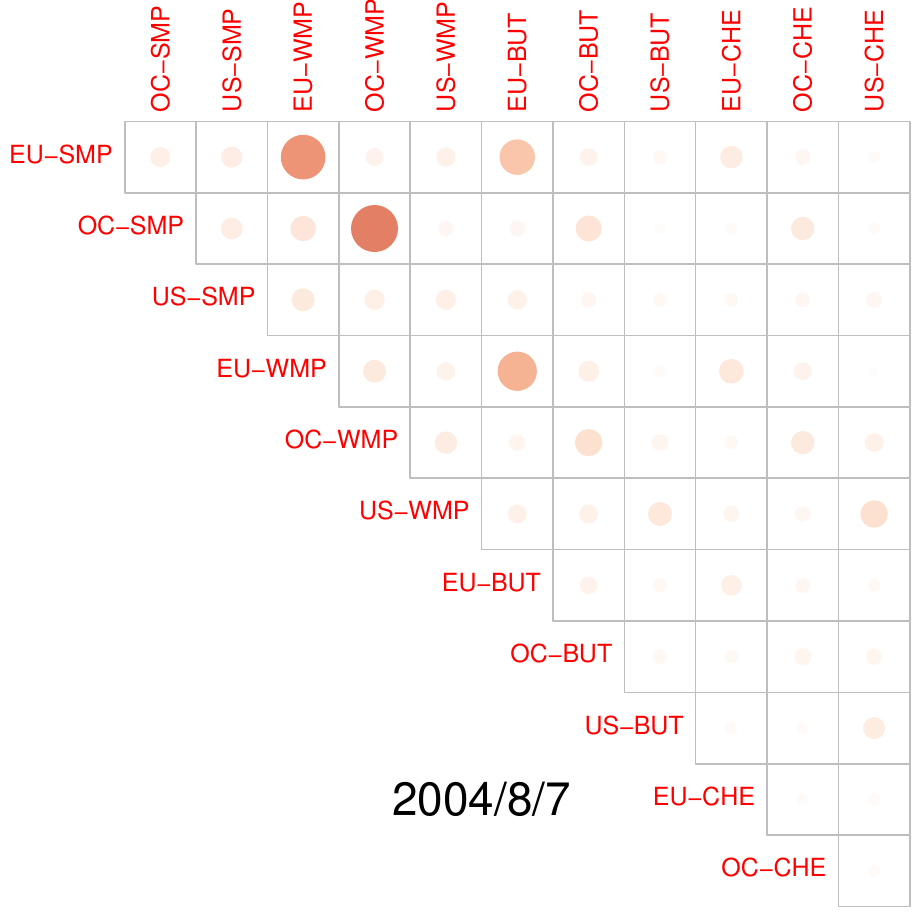}
         \caption{}
         \label{fig:b}
     \end{subfigure}
\begin{subfigure}[b]{0.193\textwidth}
         \centering
         \includegraphics[width=\textwidth, page=55]{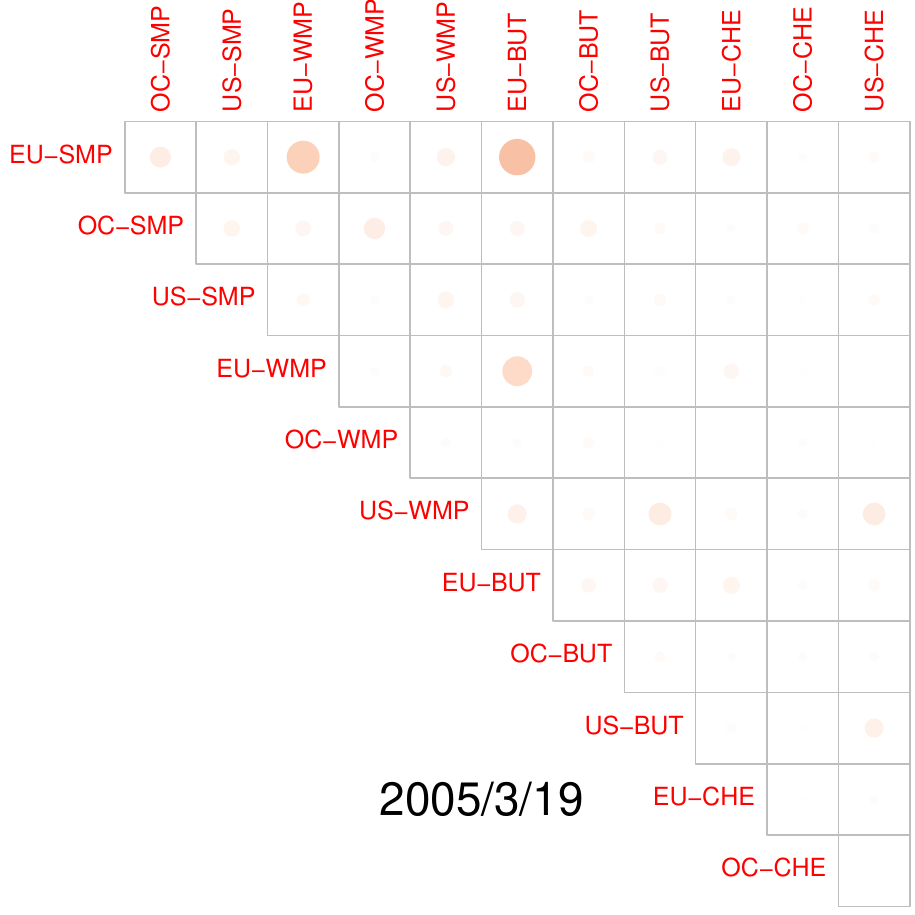}
         \caption{}
         \label{fig:c}
     \end{subfigure}
     \begin{subfigure}[b]{0.193\textwidth}
         \centering
         \includegraphics[width=\textwidth, page=71]{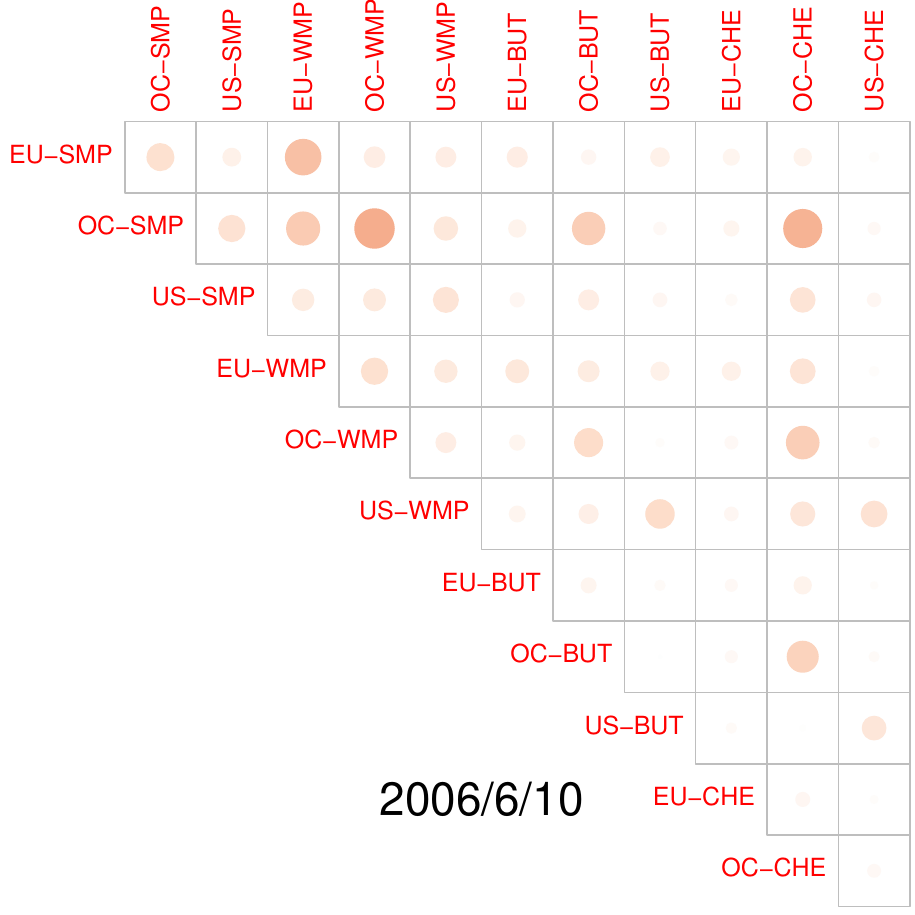}
         \caption{}
         \label{fig:d}
     \end{subfigure}
     \begin{subfigure}[b]{0.193\textwidth}
         \centering
         \includegraphics[width=\textwidth, page=82]{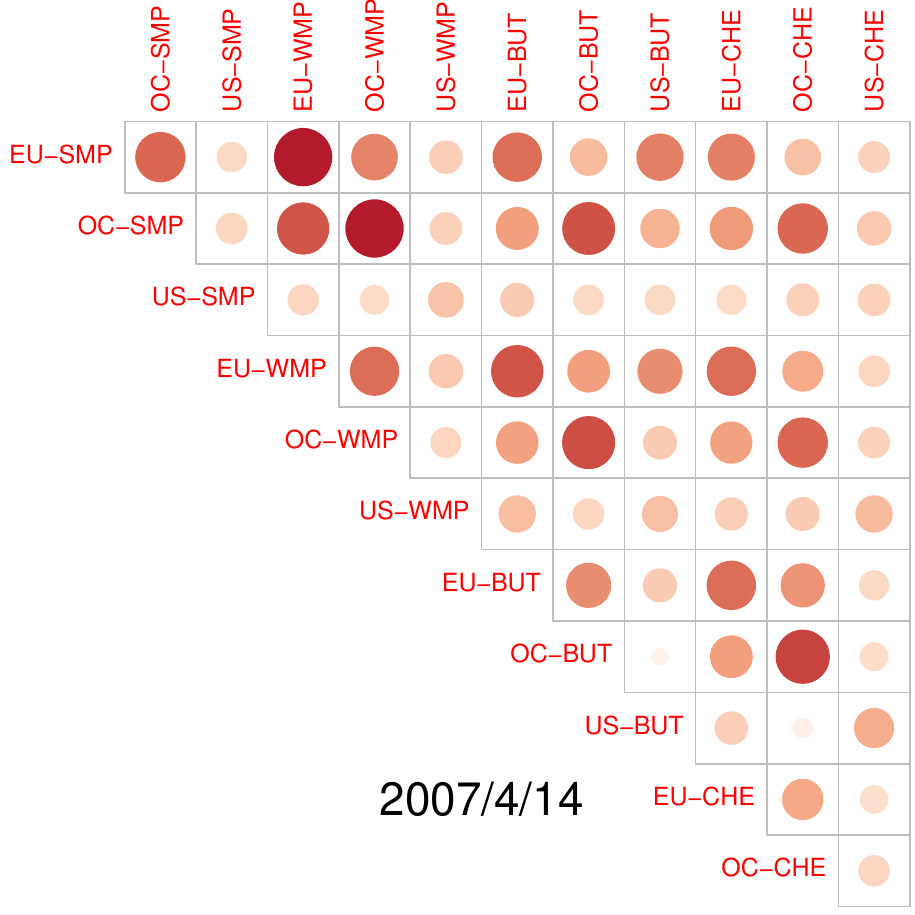}
         \caption{}
         \label{fig:e}
     \end{subfigure}
     \\[1em]
     \begin{subfigure}[b]{0.193\textwidth}
         \centering
         \includegraphics[width=\textwidth, page=83]{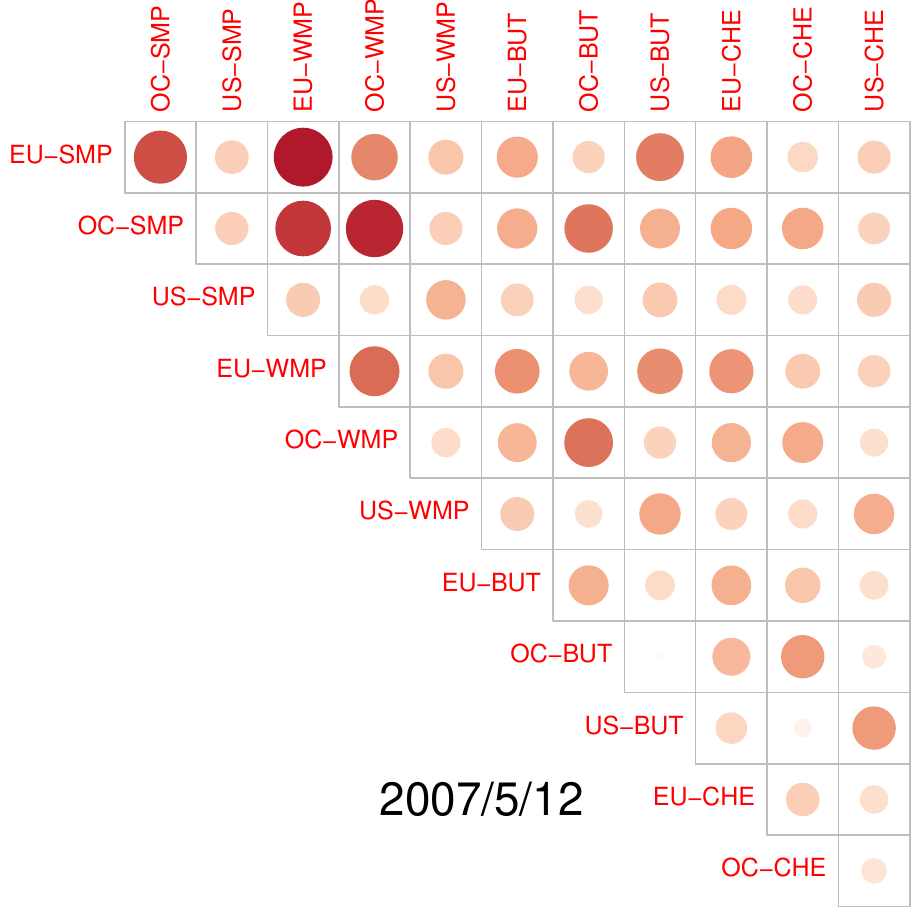}
         \caption{}
         \label{fig:f}
     \end{subfigure}
     \begin{subfigure}[b]{0.193\textwidth}
         \centering
         \includegraphics[width=\textwidth, page=100]{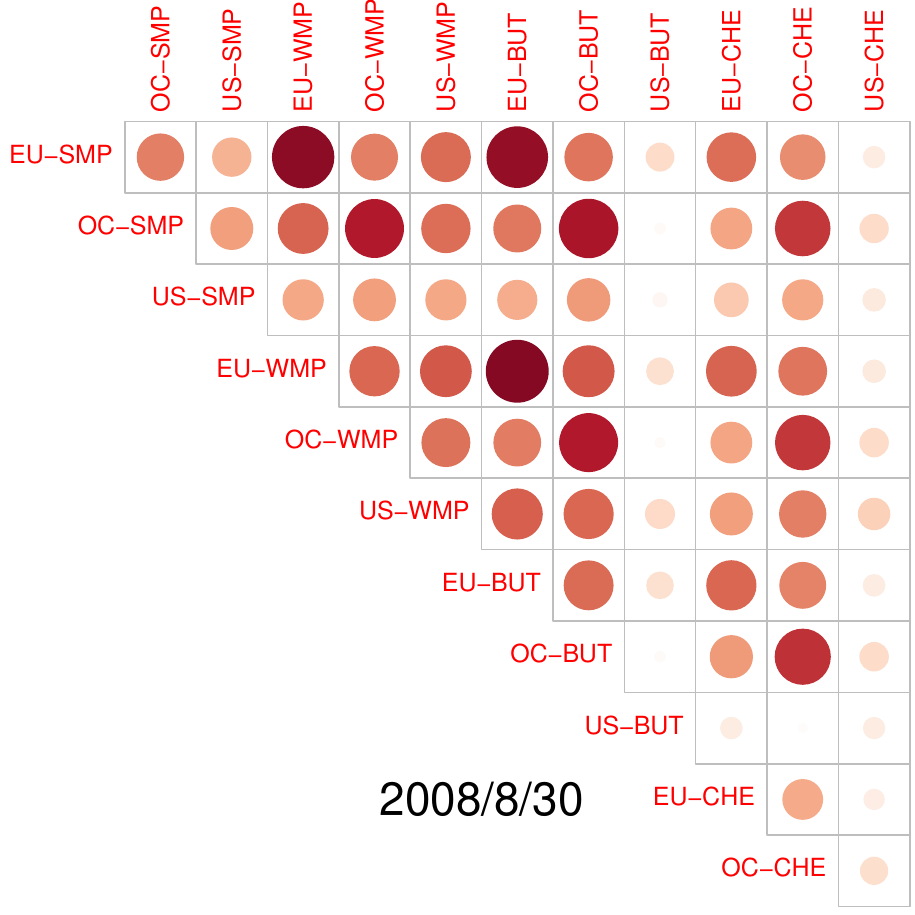}
         \caption{}
         \label{fig:g}
     \end{subfigure}
     \begin{subfigure}[b]{0.193\textwidth}
         \centering
         \includegraphics[width=\textwidth, page=178]{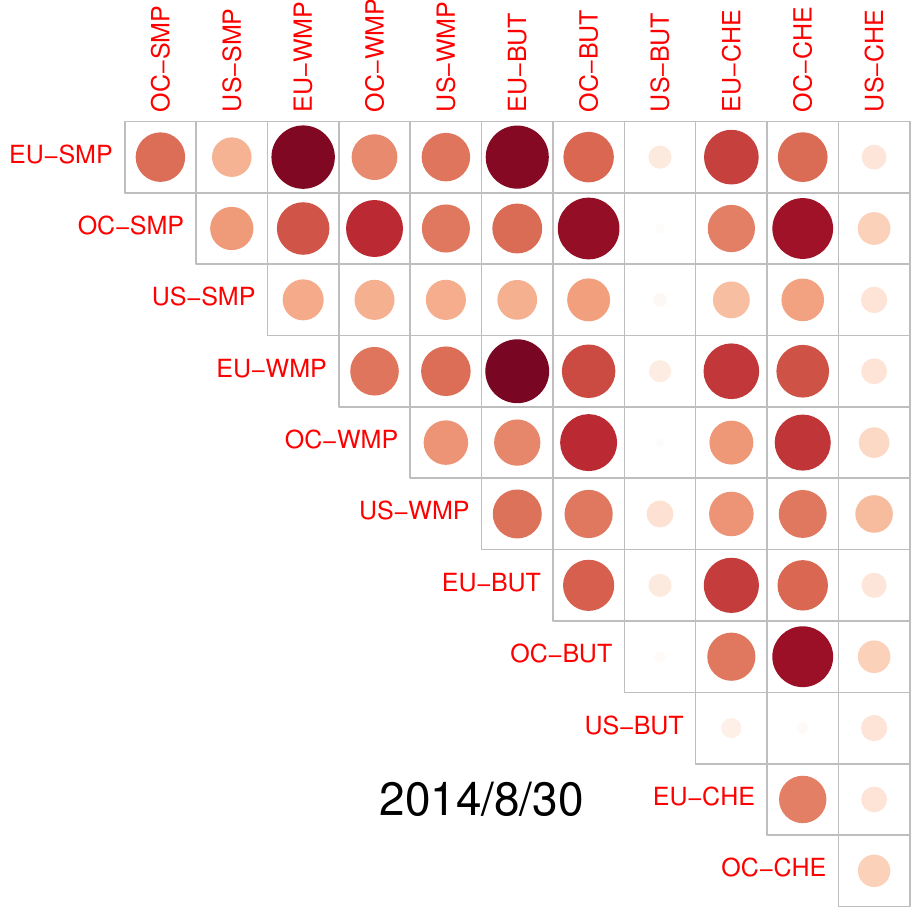}
         \caption{}
         \label{fig:h}
     \end{subfigure}
     \begin{subfigure}[b]{0.193\textwidth}
         \centering
         \includegraphics[width=\textwidth, page=190]{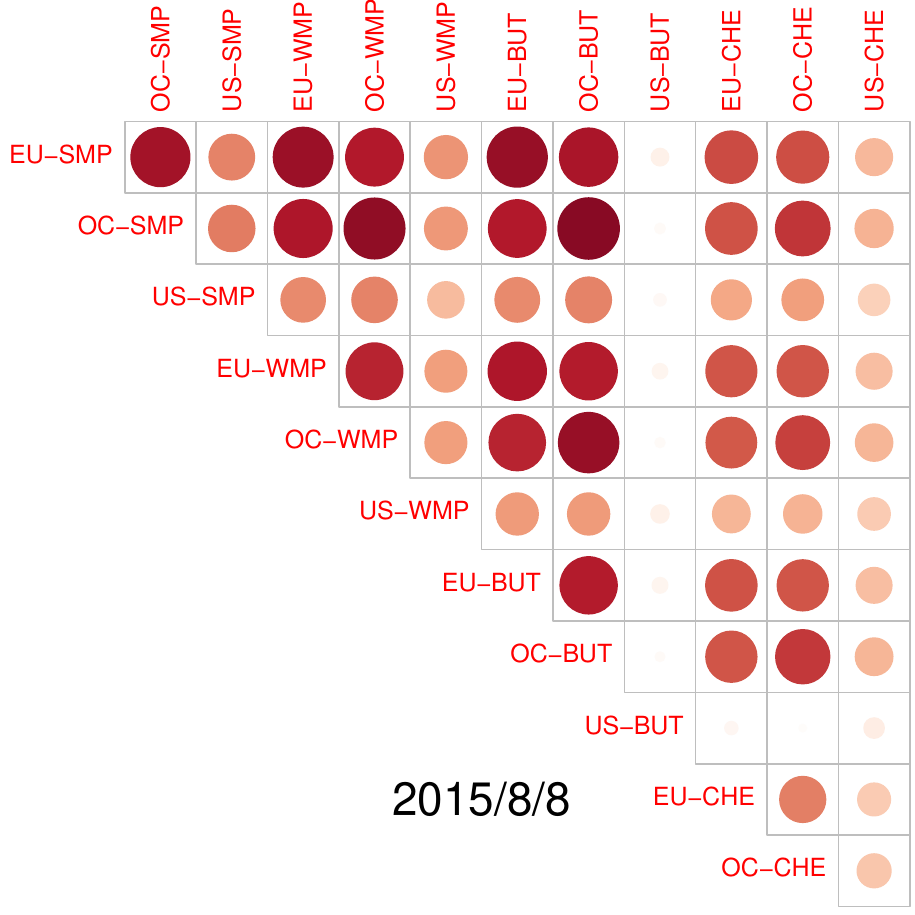}
         \caption{}
         \label{fig:i}
     \end{subfigure}
     \begin{subfigure}[b]{0.193\textwidth}
         \centering
         \includegraphics[width=\textwidth, page=204]{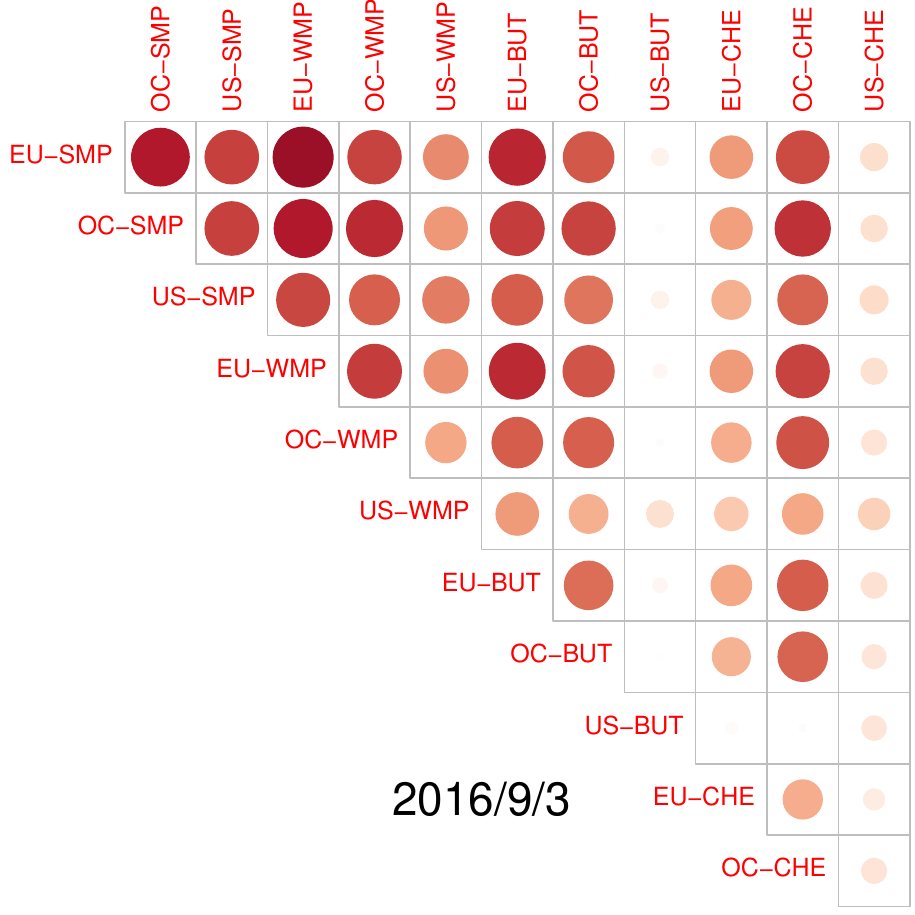}
         \caption{}
         \label{fig:j}
     \end{subfigure}
        \caption{Ten posterior mean correlation matrices}
        \label{fig:corrmatnew}
\end{figure}

Posterior means of some of the model-implied correlation matrices are displayed in Figure~\ref{fig:corrmatnew}. The matrices indicate positive and substantially time-varying correlations that tend to increase during crisis times and periods of agricultural policy changes. Subfigures \ref{fig:a} to \ref{fig:d}, corresponding to the period before the 2007 crisis year, show a low correlation between dairy commodities, while, during the 2007 crisis year (Subfigure \ref{fig:e}), the correlation increases substantially. Moreover, Subfigures \ref{fig:f} to \ref{fig:j} show a strong correlation after 2007 due to the ongoing bilateral and multilateral international trade agreements between the regions under consideration as well as to the various changes in regional agricultural policies, such as the 2008 CAP Health Check (Subfigure~\ref{fig:g}), the 2008 US Farm Act (Subfigure~\ref{fig:g}), and the 2014 CAP reform (Subfigure~\ref{fig:h}). Furthermore, in Subfigures \ref{fig:e} to \ref{fig:j}, some clusters of highly correlated dairy commodities (such as skim and whole milk powders) can be spotted, whereas cheeses show little correlation. 
Trade agreements and policy changes might have globalized the dairy market, rendering agricultural commodity prices more correlated across regions. Along these lines, the establishment of the Global Dairy Trade by Fonterra in 2008 should be mentioned; this provides electronic auctions in internationally traded commodities such as butter and milk powder, further contributing to the globalization of the international dairy market. However, as one of the reviewers of the present paper indicated, a careful inspection of Figure~\ref{fig:corrmatnew} suggests that the interplay between the domestic demand and the domestic supply for dairy commodities has a far more significant impact than that between the international demand and the international supply. This might be because the dairy markets under consideration are not perfectly integrated, as suggested by the current literature. In particular, the paper by \citet{fousekis2016price} supports the idea that the law of one price does not hold in SMP markets in the regions under consideration (the EU, Oceania, and the US), while the study by \citet{rezitis2019impact} finds that the dairy markets of the three regions (the EU, Oceania, and the US) are segmented rather than well integrated.


\subsection{VaRs and CoVaRs}
\label{sec:var}

In order to assess what our model implies in terms of risk management, we now turn towards assessing posterior quantiles of our model-implied return distribution $\bm y_t \sim \mathcal{N}_{m}(\bm 0, \bm \Sigma_t)$. More concretely, for each series $i = 1, \dots, m$ and each point in time $t$, we find the value $\text{VaR}^{t,q}_{i}$ such that
\[
P(y_{i,t} \leq \text{VaR}^{t,q}_{i}) = q, \quad q \in \{0.01, 0.05, 0.95, 0.99\}.
\]
In addition to discussing marginal VaRs, we also investigate CoVaRs \citep{adr-bru:cov}. These are defined as the quantile of some return distribution \emph{conditional} on the fact that each element of some other set of returns $\mathcal{J}$ is at their respective marginal VaR. More specifically, $\text{CoVar}^{t,q}_{i|\mathcal{J}}$ is implicitly defined by
\[
P(y_{i,t} \leq \text{CoVaR}^{t,q}_{i|\mathcal{J}} | y_{\mathcal{J},t} = \text{VaR}^{t,q}_{\mathcal{J}}) = q, \quad q \in \{0.01, 0.05, 0.95, 0.99\}.
\]

To compute CoVaR for the problem at hand, we use the fact that any multivariate normal variate can be trivially decomposed into two subsets. More concretely, we have that $\bm y_t = (\bm y_t^1, \bm y_t^2) \sim \mathcal{N}_{m}(\bm 0, \bm \Sigma_t)$ where
\[
\bm\Sigma_t = \left(\begin{array}{cc}\bm{\Sigma}_t^{11} & \bm{\Sigma}_t^{12}\\ \bm{\Sigma}_t^{21} & \bm{\Sigma}_t^{22}
\end{array}
\right).
\]
Then, the conditional distribution of $\bm{y}^1_{t}$ given known values for $\bm{y}^2_t$ is again a multivariate normal, i.e.\ $\bm y_t^1 | \bm y_t^2 \sim \mathcal{N}(\bm \mu_t^{1|2}, \bm \Sigma_t^{1|2})$ with
\begin{align} \bm\mu_t^{1|2} &= \bm\Sigma_t^{12}(\bm\Sigma_t^{22})^{-1}\bm y_t^2, \label{cond1}\\
\bm\Sigma_t^{1|2} &= \bm\Sigma_t^{11} - \bm\Sigma_t^{12} (\bm\Sigma_t^{22})^{-1} \bm\Sigma_t^{21}, \label{cond2}
\end{align}
and we can use, e.g., posterior medians of $\bm \Sigma_t$ to compute the respective CoVaRs. Note that Equations~\ref{cond1} and \ref{cond2} imply that when the two subsets under consideration are uncorrelated at a certain point in time, VaR and CoVaR coincide. On the other hand, VaR and CoVaR differ greatly whenever the correlation is high.

\begin{figure}[p]
    \centering
    \includegraphics[width=\textwidth, trim=0 0 0 12, clip]{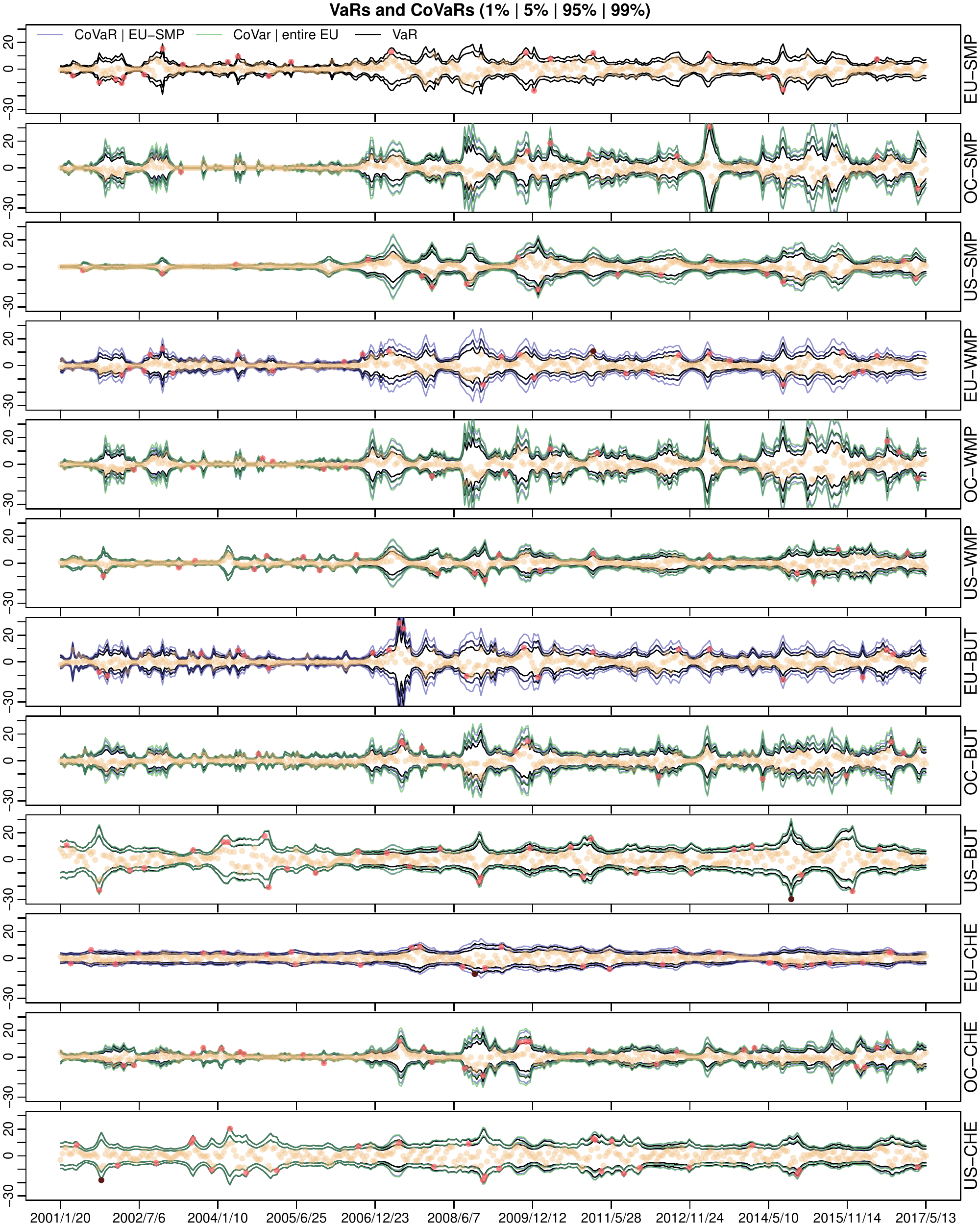}
    \caption{1\%, 5\%, 95\%, and 99\% in-sample VaRs (black lines) and CoVaRs (purple and green lines) based on the model-implied posterior median covariance matrix. Regular data points are plotted in light brown whereas VaR exceedances at the 5\%-level are plotted red and exceedances at the 1\%-level are plotted in dark brown.}
    \label{fig:var}
\end{figure}

The univariate two-sided VaRs are presented in Figure~\ref{fig:var} in black solid lines. Considering 5\% and 1\% levels, the most extreme values of the VaR estimates occur after 2006, confirming the previous findings. In particular, the most extreme value for EU-SMP occurs on 2008/12/20, while for OC-SMP on 2013/4/13, US-SMP on 2010/1/23, EU-WMP on 2009/1/10, OC-WMP on 2016/8/20, US-WMP on 2009/1/24, EU-BUT on 2007/6/9, OC-BUT on 2016/8/20, US-BUT on 2014/10/25, EU-CHE on 2009/1/10, OC-CHE on 2009/1/10, and for US-CHE on 2004/4/3. Data points are plotted as well, whereby light-brown points indicate dates that lie within the 5\% VaRs, red points indicate dates that correspond to 5\% (but not 1\%) exceedances, and dark-brown points correspond to 1\% exceedances.

In addition to standard VaRs, Figure~\ref{fig:var} also contains two CoVaR specifications. The first specification is $\text{CoVaR}_{i|1}^{t,q}$ for $i=2,\dots,12$, i.e.~the conditional quantiles of series 2 to 12 \emph{given} that EU-SMP is fixed at $\text{VaR}_1^{t,q}$. The ``CoVaR $|$ EU-SMP'' is visualized in purple and can be thought of as a measure of risk given that the returns of EU-SMP are extreme. We see that CoVaRs generally tends to be more extreme than the respective VaRs. However, note that there are certain series and certain stretches in time where VaRs and CoVaRs (almost) coincide; see, e.g., US-BUT and US-CHE. This finding again shows that shocks to a series in a certain region (here EU) do not necessarily carry over to other regions (here US).

The second CoVaR specification corresponds to ``shocking'' all four EU series at the same time: $\text{CoVaR}_{i|\{1,4,7,10\}}^{t,q}$ for $i \in \{2,3,5,6,8,9,11,12\}$. It is labeled ``CoVaR $|$ entire EU'' and depicted in green. Interestingly, the two CoVaR specifications mostly align, implying that extreme values of all EU series mean little extra risk as opposed to only seeing extreme values of EU-SMP. Only around 2007--2010, the two specifications tend to differ slightly, in particular for Oceania. Again, we see that US-BUT and US-CHE are barely affected. 

\section{Conclusions}
\label{sec:con}
This study uses a multivariate factor stochastic volatility model for estimating time-varying (dynamic) covariance and correlation matrices of dairy prices of four commodities (skim milk powder, whole milk powder, butter, and cheddar cheese) in three main regional markets (the EU, Oceania, and the US) for the period from January 20, 2001 to May 13, 2017. The model imposes a low-dimensional latent factor structure where the factors are allowed to exhibit stochastic volatility and hence determine co-movement of volatility over time. The distinguishing feature of the present study relative to most of the recent literature is that it explicitly models dairy price co-volatility linkages in the international dairy markets rather than examining the (signed) co-movements and transmissions of dairy price levels or price changes. 

The empirical results support four factors, where the first and the third factor reflect the EU and the Oceania dairy sectors, respectively. The second and fourth factors capture the milk powder markets. Note that the fourth factor expands the second factor by incorporating the US milk powder market alongside the EU and Oceania markets already included in the second factor. These results are supported by the fact that both the EU and Oceania are the main driving forces of the international dairy market, whereas the milk powders are the main dairy commodities traded by the EU, Oceania, and the US internationally. Furthermore, the empirical results show that the US is also an important participant in the global dairy market but to a lower degree than the EU and Oceania, since the US entered the international dairy market later than the other two regions.  The first and third factors show higher volatility and less persistence than the second and fourth factors. Estimated time-varying volatilities of the four factors increase after the 2006/07 global (food) crisis, which also coincides with the free trade agreements enacted from 2007 onwards and the EU and the US liberalization policy changes. These trade and policy events have created a more globalized and volatile international dairy market. However, the empirical results of the present study show that domestic demand and supply factors for dairy commodities are more significant than international demand and supply factors, indicating that international markets are not perfectly integrated. 

Marginal posterior volatilities of all dairy commodities show that the volatility of each of the dairy commodities increases after the 2006/07 global (food) crisis and for most of the commodities, and volatility rates are higher during the 2007--2010 and 2014--2016 subperiods. The Oceania dairy commodities are those showing the highest volatility. Implied correlation matrices indicate increasing correlation during the 2006/8/5 to 2007/6/9 period as well as during 2008 and 2014, which can be attributed to various regional agricultural dairy policies, such as the 2008 CAP Health Check, the 2008 US Farm Act, and the 2014 CAP reform as well as to the establishment of the Global Dairy Trade by Fonterra in 2008. The most extreme values of the VaR estimates for most dairy commodities occur after the year 2006 and are in line with the general volatility findings of the present study. Finally, CoVaRs conditional on extreme values in the EU tend to be higher than standard VaRs, in particular in times of high correlations.

Both publicly as well as privately, a broad range of instruments should be utilized to manage price volatility. Such instruments include over-the-counter contracts, forward contracting, futures contracts, and insurance contracts. Several private and public sector risk management instruments have already been employed by the US dairy sector, which could be used as indicators for other dairy sectors such as by the EU dairy sector. Finally, it is recommended that some public measures such as counter-cyclical measures are maintained in order to mitigate the effects of high price volatility. However, such instruments should not inhibit the development of private measures. 

\section*{Data Availability Statement}
The data that support the findings of this study are available from the corresponding author upon reasonable request.

\onehalfspacing
\printbibliography

\end{document}